\documentclass[a4paper,fleqn,usenatbib]{mnras}

\usepackage{newtxtext,newtxmath}
\usepackage[T1]{fontenc}
\usepackage{ae,aecompl}

\usepackage{tabularx}	% Setting table sizes
\usepackage{graphicx}	% Including figure files
\usepackage{amsmath}	% Advanced maths commands
\usepackage{amssymb}	% Extra maths symbols
\defcitealias{Ser16}{Paper I}

\title[Spectroscopy of double quasar candidates]{Spectroscopic follow-up of 
double quasar candidates} 

\author[V. N. Shalyapin et al.]{
V. N. Shalyapin,$^{1,2}$\thanks{E-mail: vshal@ukr.net}
A. V. Sergeyev,$^{3,4}$ 
L. J. Goicoechea$^{2}$
and A. P. Zheleznyak$^{4}$
\\
$^{1}$O. Ya. Usikov Institute for Radiophysics and Electronics, National Academy 
of Sciences of Ukraine, 12 Ac. Proskura St., 61085 Kharkov, Ukraine\\ 
$^{2}$Departamento de F\'isica Moderna, Universidad de Cantabria, Avda. de Los
Castros s/n, 39005 Santander, Spain\\
$^{3}$Institute of Radio Astronomy of the National Academy of Sciences of 
Ukraine, Mystetstv 4, 61002 Kharkiv, Ukraine\\
$^{4}$Institute of Astronomy of V. N. Karazin Kharkiv National University, 
Svobody Sq. 4, 61022 Kharkiv, Ukraine
}

\date{Accepted XXX. Received YYY; in original form ZZZ}

\pubyear{2017}

\begin{document}
\label{firstpage}
\pagerange{\pageref{firstpage}--\pageref{lastpage}}
\maketitle

% Abstract of the paper
\begin{abstract}
We report the results of an optical spectroscopic follow-up of four double quasar 
candidates in the Sloan Digital Sky Survey (SDSS) database. SDSS J1617+3827 is 
most likely a lensed quasar at $z$ = 2.079, consisting of two images with $r 
\sim$ 19$-$21 and separated by $\Delta \theta \sim$ 2\arcsec. We identify an 
extended source northeast of the brightest image as an early-type lensing galaxy 
at $z$ = 0.602, and detect a candidate for the main deflector in the vicinity of 
the faintest image. SDSS J2153+2732 consists of two distinct but physically 
associated quasars at $z \sim$ 2.24 ($r \sim$ 19$-$20, $\Delta \theta$ = 
3\farcs6). Although this system might be a binary quasar, there is evidence of a 
collision or merger within a galaxy cluster at an early stage. The other two 
candidates are projected pairs of active galactic nuclei: SDSS J1642+3200 
($\Delta \theta \sim$ 3\arcsec) comprises a distant quasar ($r \sim$ 18) at $z$ = 
2.263 and the active nucleus ($r >$ 20) of a galaxy at $z \sim$ 0.3, while SDSS 
J0240$-$0208 ($r \sim$ 18$-$19, $\Delta \theta \sim$ 1\arcsec) is a pair of 
quasars at $z$ = 1.687 and $z$ = 1.059. In each of these two systems, the 
background quasar only suffers a weak gravitational lensing effect by the host 
galaxy of the foreground active nucleus, so the host galaxy mass is constrained 
to be less than (2.9$-$3) $\times$ 10$^{11}$ M$_{\odot}$ inside 10 kpc.
\end{abstract}

% Select between one and six entries from the list of approved keywords.
% Don't make up new ones.

\section{Introduction}
\label{sec:intro} 

Quasars that are being strongly lensed by gravitational fields of foreground 
massive galaxies serve as unique probes of our Universe 
\citep[e.g.][]{Sch06,Tre10,Jac12}. For instance, observations of these 
multiply-imaged quasars are used to constrain cosmological parameters and sizes 
of quasar accretion discs. Multiple quasar systems are also valuable tools for 
investigating the structure and evolution of lensing galaxies. In addition, 
physically associated, small-separation quasar pairs lead to relevant information 
on small-scale quasar clustering, providing evidence that direct interactions 
trigger or enhance quasar activitity \citep[e.g.][]{Hen06a,Kay12}. Although such 
systems are generically termed binary quasars, here we distinguish between two
different scenarios \citep[e.g.][]{Djo87}: the term binary quasar describes two 
quasars residing at the nuclei of two galaxies that are gravitationally bound, 
while we refer to close quasar pair in a cluster as the active nuclei of two 
colliding/merging galaxies in a high density region. Projected quasar pairs can 
also be used to study environments of foreground quasars 
\citep[e.g.][]{Hen06b,Lau18} or constrain their masses \citep[e.g.][]{Cla00}.

The Sloan Digital Sky Survey (SDSS) database\footnote{\url{http://www.sdss.org/}}
is being deeply mined to discover new multiply-imaged, gravitationally lensed 
quasars \citep[e.g.][]{Ina12,Mor16}, and the identification of new binary quasars 
and other interesting systems (see above) is a by-product of these efforts. 
Searching for new doubly-imaged quasars in the SDSS-III DR10 \citep{Ahn14,Par14}, 
we selected a subsample of 14 double quasar candidates (SS2), where each of the 
targets in this subsample consisted of a confirmed quasar and an unidentified 
point-like source \citep[see Table 1 of][henceforth Paper I]{Ser16}. To 
construct SS2, we checked for the presence of some point-like source in the 
vicinity ($\leq$ 6\arcsec\ apart) of each DR10Q quasar \citep{Par14} in the range 
RA = 180$-$360\degr, at redshift 1 $< z <$ 5 and brighter than $r$ = 20. For each 
widely-separated quasar-companion pair ($\Delta \theta >$ 2\arcsec), we also 
compared the $u-g$, $g-r$, $r-i$ and $i-z$ colours of the quasar and its 
companion. After visual inspection, we finally selected the 14 promising 
candidates. About 80\% of the SS2 quasars span the redshift interval $z$ = 
2$-$2.6, and most SS2 targets are widely-separated pairs.

Complementary deep-imaging of 13 SS2 candidates at the Maidanak Astronomical 
Observatory\footnote{\url{http://www.maidanak.uz/}} (MAO) was used to select 
three superb candidates, i.e., quasar-companion pairs showing parallel flux 
variations on a 10-year timescale and evidence for the presence of a lensing 
galaxy (extended residual source). In \citetalias{Ser16}, one out of these three 
golden targets was confirmed as the optically bright, wide separation double 
quasar SDSS J1442+4055 at redshift $z$ = 2.575 \citep[see also][]{Mor16}, while 
the other two superb candidates SDSS J1617+3827 and SDSS J1642+3200 were not 
spectroscopically observed. Here, we present spectra of these two last systems 
that allow us to discuss their nature. For SDSS J1617+3827, in addition to 
the extraction of spectra, we carefully analyse its best acquisition frame. In 
this paper, we also describe 
spectroscopic observations of another target in the SS2 subsample (SDSS 
J2153+2732) and the pseudo-galaxy SDSS J0240-0208. SDSS J2153+2732 is a 
quasar-companion pair showing anti-parallel flux variations and no evidence for a 
lensing galaxy. Although SDSS J0240-0208 was identified as a galaxy in the SDSS 
database, we serendipitously discovered that this target consists of two 
unidentified point-like sources separated by $\sim$ 1\arcsec. The spectra of SDSS 
J2153+2732 and SDSS J0240-0208 are used to identify all main sources in these 
systems. 

\section{Two superb candidates}
\label{sec:golden}

\subsection{SDSS J1617+3827} 
\label{sec:q1617}

\begin{figure}
\centering
\includegraphics[width=\columnwidth]{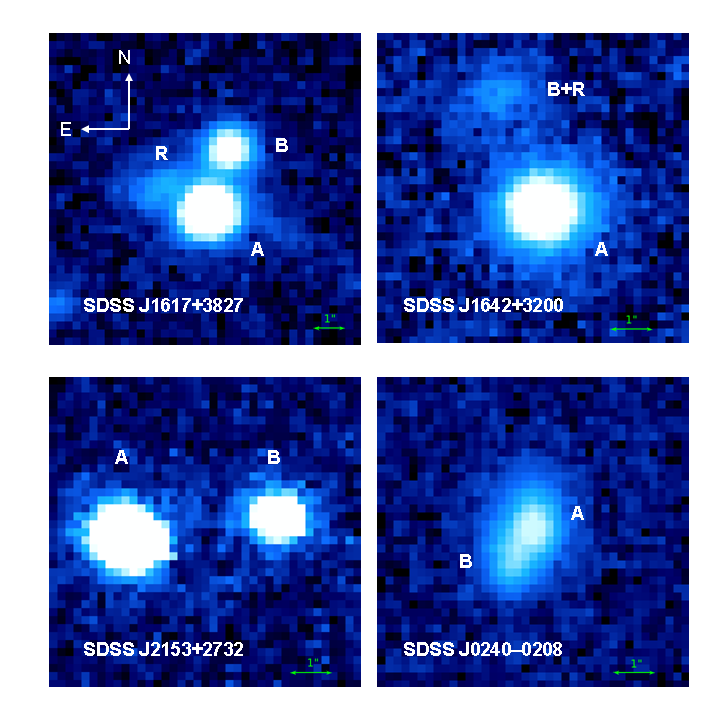}
\caption{Acquisition frames obtained just before spectroscopic observations. The 
40-s $r$-band exposure of SDSS J1617+3827 was taken with the GTC (before using 
the AR configuration; top left panel), whereas the other three frames correspond 
to short white light exposures with the NOT.}
\label{fig:acquis}
\end{figure}

SDSS J1617+3827 consists of a quasar at $z$ = 2.079 with $r \sim$ 19 (A), its 
companion with $r \sim$ 21 (B) and a residual source (R) located northeast of A
(see the top left panel of Fig.~\ref{fig:acquis} and Fig. 2 in 
\citetalias{Ser16}). The SDSS position of A is RA (J2000) = 244\fdg47052 and Dec. 
(J2000) = +38\fdg46030, and A and B are separated by 2\farcs1. This target was 
initially observed in 2016 May using the SPRAT spectrograph on the 2.0m Liverpool 
Telescope. Despite taking a 5000-s total exposure, the data did not confirm the
similarity between the spectra of A and B. The SPRAT spectrum of B seems to 
include emission lines similar to those of A, but the signal is relatively poor 
and it could be dominated by residual light of the brighter component. Due to the
ambiguity of these initial results, SDSS J1617+3827 was also observed in dark 
time with the OSIRIS instrument on the 10.4m Gran Telescopio Canarias (GTC). On 
2017 March 1, using the grism R500B and the 1\farcs23-width slit aligned with the 
axis defined by A and B, we took 3$\times$900 s exposures with FWHM seeing of 
1\farcs03, 0\farcs89 and 0\farcs87. In a second observing run during such night, 
after aligning the slit to the direction joining A and R, we obtained three 
additional 900 s exposures with 0\farcs76 FWHM seeing.  

To extract the spectrum of each individual source, we followed a technique 
similar to that of \citet{Sha14} and \citet{Goi16}. We used the astro-photometric 
solution from the acquisition frames in the $r$ band to make two ideal 2D 
light models with two on-axis point-like sources: AB for the first slit 
orientation and AR for the second one. These ideal models were then convolved 
with 2D Moffat PSFs having a power index $\beta$ = 2.5, masked with the slit 
transmission and integrated across the slit. Thus, for each spectroscopic 
exposure, our 1D model at each wavelength bin included four free parameters in a 
first iteration, i.e., the position of A, the width of the Moffat function, and 
the total fluxes of A and the other source (B or R). The two position-structure 
parameters were then fitted to smooth polynomial functions of the observed 
wavelength, leaving only the total fluxes as free parameters in a second 
iteration. The initial configuration led to flux-calibrated spectra $A1$ and $B$, 
whereas the second configuration yielded flux-calibrated spectra $A2$ and $R$. 

\begin{figure}
\centering
\includegraphics[width=\columnwidth]{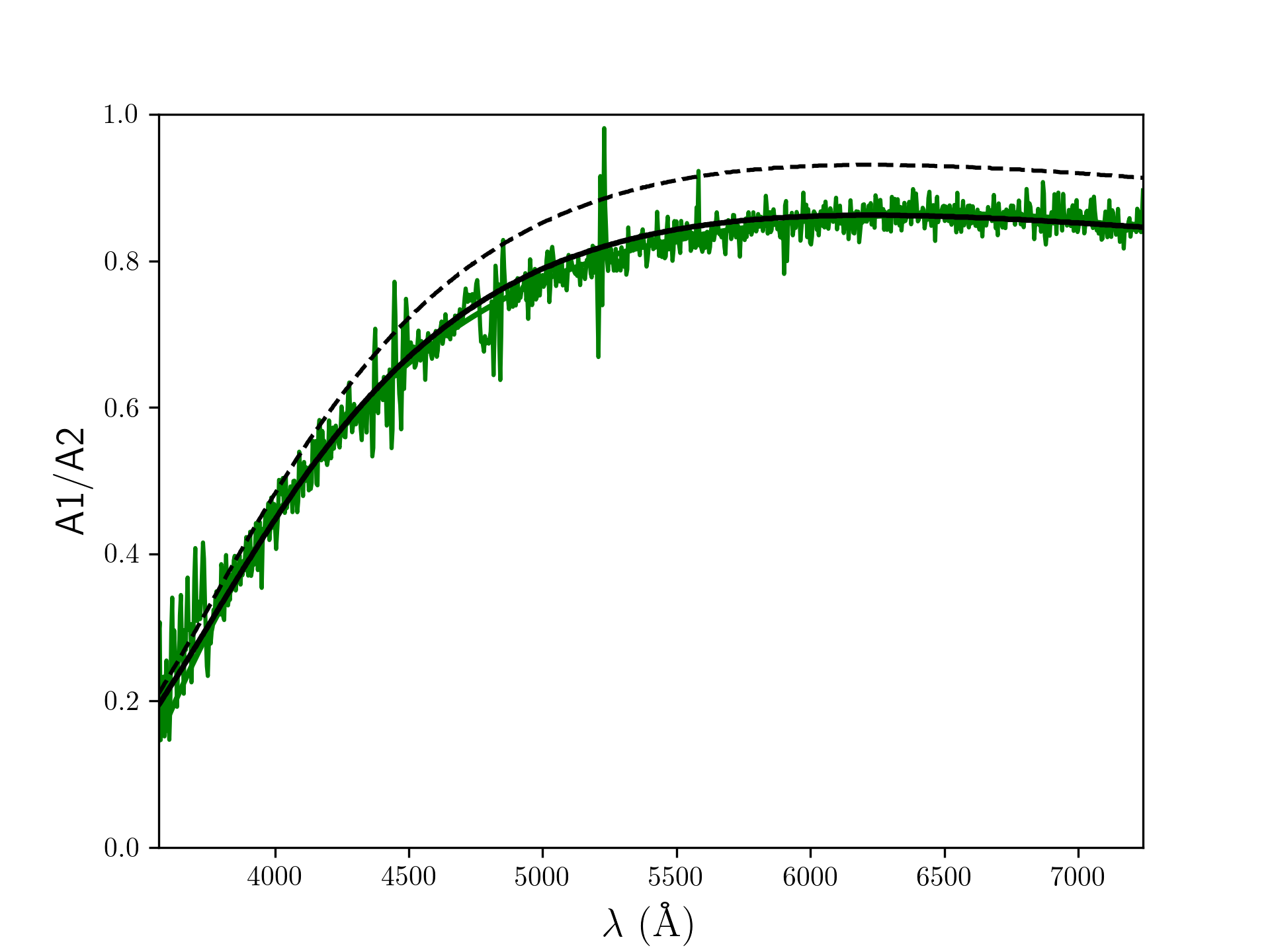}
\caption{GTC spectral ratio for the A component in SDSS J1617+3827. The green 
lines show the ratio between measured spectra of A for the two slit orientations. 
We also display expected spectral ratios when DAR (black dashed line) and DAR 
plus an achromatic multiplicative factor (black solid line) are considered.}
\label{fig:GTCq1617Ar}
\end{figure}

The spectral ratio $A1/A2$ strongly depends on wavelength (see green lines in 
Fig.~\ref{fig:GTCq1617Ar}), suggesting that differential atmospheric refraction 
(DAR) is playing a relevant role and on-axis sources throughout the entire 
spectral range is not a realistic approach. Hence, to account for DAR-induced 
spectral distortions, we considered on-axis sources only at 6225 \AA, since the 
adquisition frame was taken in the SDSS $r$ passband. The DAR produces chromatic 
offsets of sources across the slit, and thus wavelength-dependent slit losses 
\citep{Fil82}. Taking into account the airmass and the slit position angle for 
each exposure, we derived the slit losses for both configurations AB and AR, as 
well the expected spectral ratio $A1/A2$ (see the black dashed line in 
Fig.~\ref{fig:GTCq1617Ar}). The expected ratio $A1/A2$ follows the same trend as 
the measured one, but having a larger amplitude. Multiplying it by an achromatic 
factor of 0.926, we reproduce the measured spectral ratio (see the black solid 
line in Fig.~\ref{fig:GTCq1617Ar}). This last factor is likely related to 
changes in atmospheric transparency throughout the observing night.

\begin{figure}
\includegraphics[width=\columnwidth]{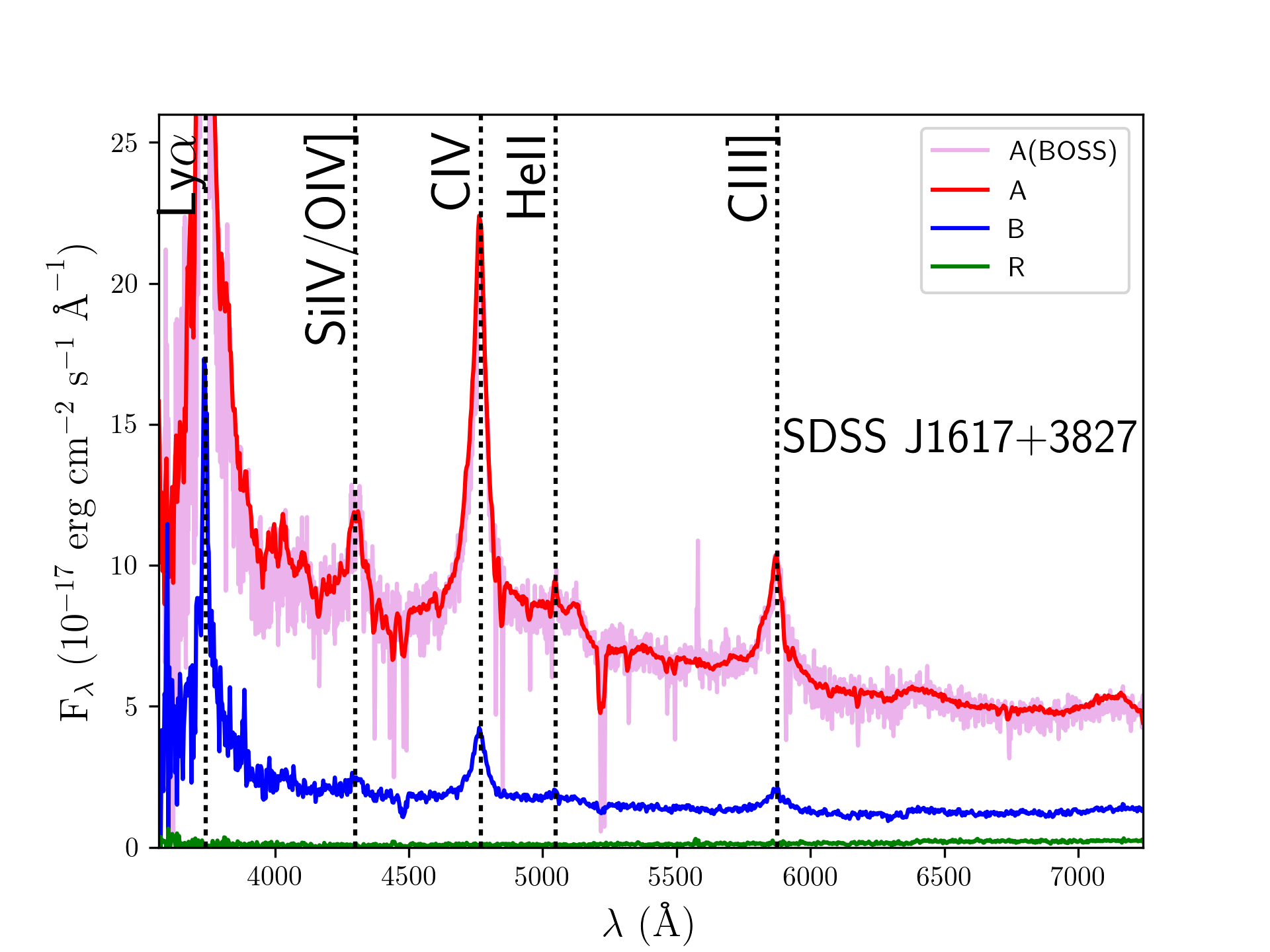}
\includegraphics[width=\columnwidth]{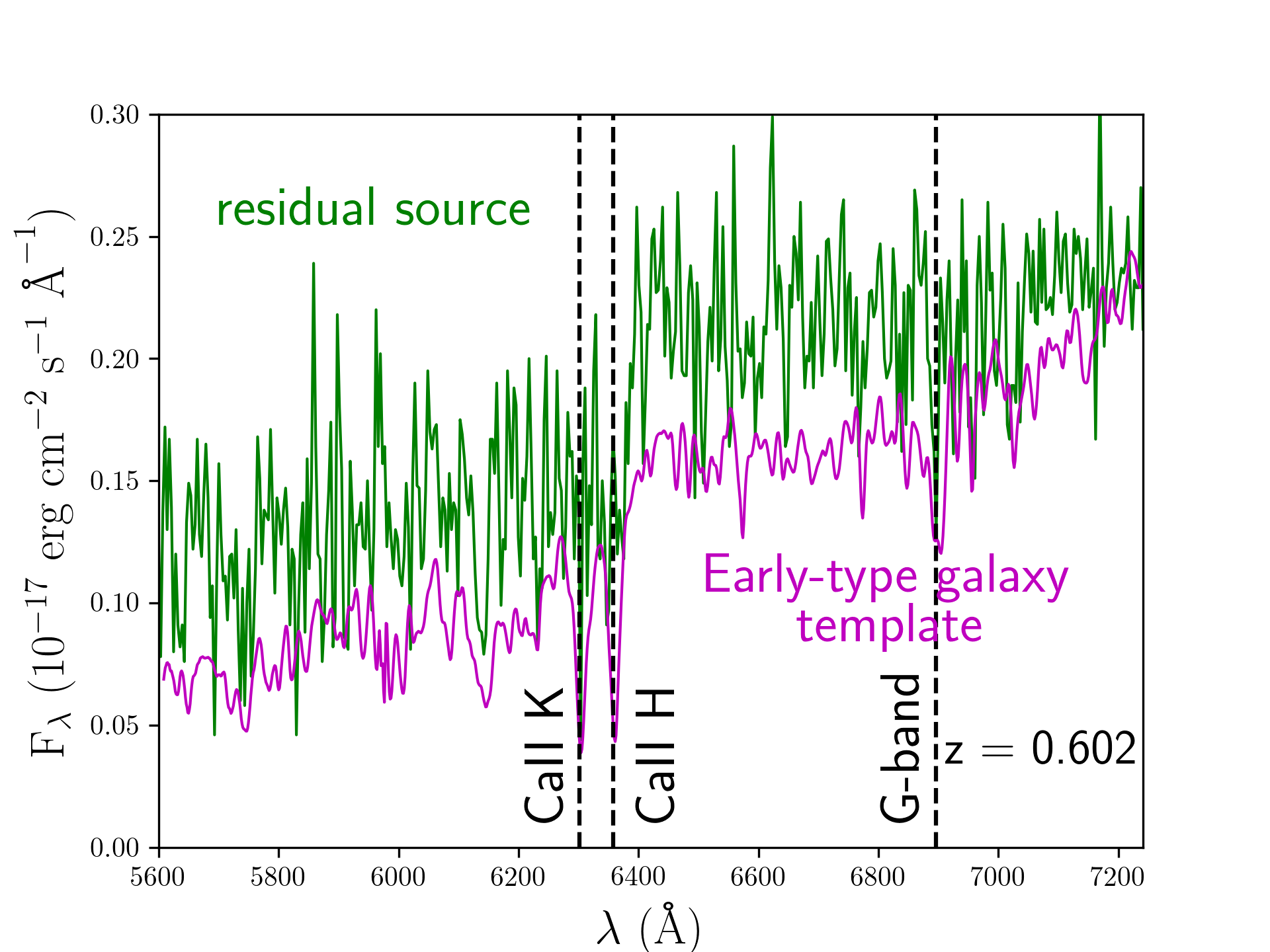}
\caption{Spatially resolved spectroscopy of SDSS J1617+3827 from GTC 
observations. Top: spectra of the confirmed quasar A (red line), its companion B 
(blue line) and the residual source R (green line). We also display the SDSS-BOSS 
spectrum of A (magenta line) for comparison purposes. Vertical dotted lines 
indicate emission lines at $z$ = 2.079. Bottom: a zoomed-in version of the 
reddest part of the R spectrum (green line) along with the red-shifted spectral 
template of an early-type galaxy ($z$ = 0.602; magenta line). Vertical dashed 
lines are associated with absorption features.}
\label{fig:GTCq1617ABR}
\end{figure}

After correcting for slit losses in $A1$, $A2$, $B$ and $R$, and combining $A1$ 
and $A2$, we obtained final spectra of A, B and R. Table~\ref{tab:q1617dat} 
includes wavelengths (Col. 1), and fluxes of A, B and R (Cols. 2, 3 and 4). The 
GTC spectra are also shown in Fig.~\ref{fig:GTCq1617ABR}. In the top panel of 
Fig.~\ref{fig:GTCq1617ABR}, our A spectrum is compared with the SDSS-BOSS 
spectrum of this component, as well as the spectra of B and R. As expected for a 
double quasar, the spectra of A and B show identical emission-line redshifts. The 
bottom panel of Fig.~\ref{fig:GTCq1617ABR} incorporates a zoomed-in version of 
the spectrum of R at the longest wavelengths. Despite the faintness of this 
residual source, we identify it as an early-type lensing galaxy at $z$ = 0.602. 

\begin{table}
\centering
\caption{GTC spectra of the three components in SDSS J1617+3827. Observed 
wavelengths ($\lambda$ values) are in \AA\ and fluxes ($F_{\lambda}$ values) are 
in 10$^{-17}$ erg cm$^{-2}$ s$^{-1}$ \AA$^{-1}$. The full table is available 
online in a machine-readable ASCII format. A portion is shown here for guidance 
regarding its form and content.}
\label{tab:q1617dat}
\begin{tabular}{cccc} 
\hline
$\lambda$ & $F_{\lambda}$(A) & $F_{\lambda}$(B) & $F_{\lambda}$(R) \\
\hline
3566.450 & 15.814 & 0.948 &  0.138 \\
3570.042 & 14.197 & 1.827 & -0.275 \\
3573.634 & 10.100 & 0.000 &  0.370 \\
3577.226 & 12.591 & 4.153 &  0.157 \\
3580.819 &  9.798 & 2.793 &  0.253 \\
3584.411 & 10.852 & 1.992 &  0.200 \\
3588.003 &  8.625 & 5.442 &  0.224 \\
3591.595 & 12.829 & 5.066 &  0.189 \\
\hline
\end{tabular}
\end{table}

\begin{table*}
\centering
\caption{Astro-photometric solution for SDSS J1617+3827 in the $r$ band. The A 
image is at the origin of coordinates, and the positive directions of $x$ and $y$
are defined by west and north, respectively. For each extended object (de 
Vaucouleurs profile), in addition to its position and magnitude, we give the 
effective radius $R_{\rm{eff}}$, the ellipticity $e$, and the position angle of 
the major axis $\theta_e$ (it is measured east of north).}
\label{tab:galfitsol}
\begin{tabular}{lcccccc} 
\hline
Object & $x$ (\arcsec) & $y$ (\arcsec) & $r$ (mag) & $R_{\rm{eff}}$ (\arcsec) & 
$e$ & $\theta_e$ (\degr)\\
\hline
A & 0 & 0 & 19.37 $\pm$ 0.01 & --- & --- & ---\\ 
B & 0.643 $\pm$ 0.003 & 1.983 $\pm$ 0.002 & 21.08 $\pm$ 0.01 & --- & --- & ---\\ 
R & $-$1.509 $\pm$ 0.018 & 0.896 $\pm$ 0.012 & 21.01 $\pm$ 0.16 & 3.15 $\pm$ 0.55 & 
0.32 $\pm$ 0.05 & $-$68.7 $\pm$ 6.1\\ 
R2 & 0.379 $\pm$ 0.041 & 1.689 $\pm$ 0.061 & 22.00 $\pm$ 0.21 & 1.95 $\pm$ 0.63 & 
0.34 $\pm$ 0.11 & $-$5.1 $\pm$ 12.9\\ 
\hline
\end{tabular}
\end{table*}

The high-resolution SDSS-BOSS spectrum of A, taken on 2012 May 20, and the 
{\small IRAF} task {\small SPLOT}\footnote{IRAF is distributed by the National 
Optical Astronomy Observatories, which are operated by the Association of 
Universities for Research in Astronomy, Inc., under cooperative agreement with 
the National Science Foundation. The SPLOT task is used to plot and analyse 
spectra} were used to search for Fe/Mg absorption systems, and accurately measure 
redshifts and line strengths of \ion{Mg}{ii} $\lambda\lambda$ 2796, 2803 
absorption doublets. We found four different absorbers at $z$ = 0.6020, 0.8656, 
1.1129 and 1.5927, the second being the strongest one. These results 
basically agree with previous measures that were based on an early SDSS spectrum 
in 2003 April. \citet{Pro06} detected the two first absorbers when analysing 
\ion{Mg}{ii} $\lambda\lambda$ 2796 lines having a rest-frame equivalent width 
($EW$) exceeding 1 \AA. \citet{Law12} and \citet{Sey13} also reported $EW$ values
for the strongest \ion{Mg}{ii} $\lambda\lambda$ 2796, 2803 doublet lines at $z$ = 
0.8656. For example, \citet{Sey13} estimated $EW \sim$ 2.4 \AA\ (2796 \AA) and 2 
\AA\ (2803 \AA), which roughly match our results using the spectrum more recent 
and less noisy: $EW$ = 2.21 \AA\ (2796 \AA) and 2.25 \AA\ (2803 \AA). 

The GTC data have worse spectral resolution than SDSS-BOSS ones, but the 
signal-to-noise ratio is higher and the absorption in both quasar images can be 
compared with each other. Regarding the nearest intervening system ($z$ = 0.6020)
that is likely associated with the lensing galaxy northeast of A, the $EW$ of 
the \ion{Mg}{ii} doublet in B is large compared with that in A. There is a 
difference in a factor $>$ 2 between the two images. However, for the second 
system at $z$ = 0.8656, we estimated $EW$(A) $\sim$ 1.5 $EW$(B). Unfortunately, 
the position in the sky of such absorber is unknown, so it is hard to discuss the 
role that its gravitational field plays. The lensed components (A and B) and 
the lensing galaxy R are not aligned, and we could not make reasonable strong 
lens models using exclusively data of A, B and R as constraints. Therefore, the 
residual object in the top left panel of Fig.~\ref{fig:acquis} is not the main 
lensing galaxy in this putative lens system. 

\begin{figure}
\centering
\includegraphics[width=\columnwidth]{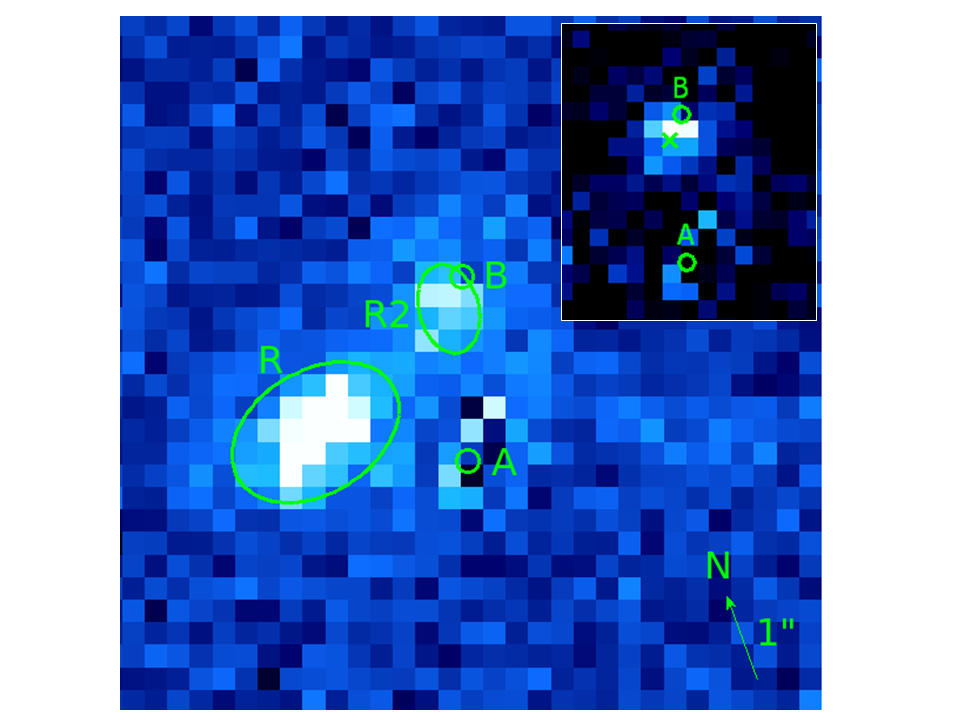}
\caption{{\small GALFIT} modelling results of a square subframe around the lens 
system with 32 pixels per side. The model includes the sources A, B, R and R2 
(see main text). We show residual fluxes after subtracting only the quasar images 
A and B (small circles). The positions of R and R2 correspond to the centres of 
the two ellipses, which have the values of $e$ and $\theta_e$ in 
Table~\ref{tab:galfitsol}, and major axes proportional to the $R_{\rm{eff}}$ 
values. The positive signal after subtracting A, B and R is displayed in the top 
right subpanel. Note position of R2 (cross) close to the faintest image B.}
\label{fig:galfit}
\end{figure}

To confirm the strong gravitational lens nature of SDSS J1617+3827, we 
searched for the elusive primary deflector in the best adquisition frame (see 
Fig.~\ref{fig:acquis}). Our aim was to detect an additional residual source R2 
and find out its properties. Despite this GTC $r$-band exposure lasted only 40 s, 
it is equivalent to a relatively deep exposure of $\sim$ 1080 s with a 2m 
telescope. The FWHM seeing was stable at around 0\farcs75. The system 
subframe consisted of 32$\times$32 pixels (with pixel scale and position angle of 
0\farcs2544 and $-$20\degr, respectively), and the objects A and B were described 
as two stellar-like sources, i.e., using the PSF of a nearby star, while each 
residual object was modelled as a de Vaucouleurs profile convolved with the PSF. 
We used the {\small GALFIT} software \citep{Pen02,Pen10} to initially model the 
system as a superposition of A, B and R. In a subsequent fit, we considered a 
fourth source R2. When introducing R2, the number of degrees of freedom is 
reduced by 6, but the $\chi^2$ value decreases by 63. Thus, we detect a faint 
source R2 close to B and aligned with both quasar images (see the top right 
subpanel in Fig.~\ref{fig:galfit}), which provides evidence supporting that SDSS 
J1617+3827 is a lensed quasar. 

The {\small GALFIT} solution for the A+B+R+R2 model is shown in 
Table~\ref{tab:galfitsol}. This solution has a reduced $\chi^2$ value of 0.88, 
and incorporates positions and structure parameters for the light distributions 
of R (secondary lensing object) and R2 (candidate for the main lensing object). 
In Fig.~\ref{fig:galfit}, we display the residuals after subtracting the 
modelling results for A and B, as well as the positive residuals after 
subtracting the models of A, B and R (top right subpanel). New deep IR imaging is 
required to accurately determine astro-photometric parameters of R and R2. 
Additionally, an accurate reconstruction of the lensing mass distribution for 
SDSS J1617+3827 is not yet feasible.  
 
\subsection{SDSS J1642+3200} 
\label{sec:q1642} 

SDSS J1642+3200 consists of a quasar at $z$ = 2.263 with $r \sim$ 18 (A), its 
companion with $r >$ 20 (B) and a residual source (R) around the fainter object.
Although B+R appears as a dim light distribution in the top right panel of 
Fig.~\ref{fig:acquis}, we were able to resolve both sources from the deep 
$r$-band exposure at the MAO (see Fig. 2 in \citetalias{Ser16}). Component A
is located at RA (J2000) = 250\fdg71282 and Dec. (J2000) = +32\fdg00811, and A 
and B are 2\farcs9 apart. The source R could be physically associated with B, so 
B would be the central region of a galaxy. In fact, the SDSS database clasifies 
B+R as a galaxy with photometric redshift of 0.25 $\pm$ 0.07. However, even in 
this case, we would not dealing with a normal galactic nucleus because B can be 
resolved as a point-like source and shows optical variability. To clarify the 
true nature of the system, SDSS J1642+3200 was spectroscopically observed on 2016 
March 3 using the ALFOSC instrument on the 2.5m Nordic Optical Telescope (NOT). 
We also used the grism \#18 and the 1\farcs0-width slit, putting the long slit in 
the direction joining A and B. Four consecutive exposures of 1000 s each were 
done under typical seeing conditions.  
    
As the A and B components are separated by $\Delta \theta \sim$ 3\arcsec, 
in each individual frame, we clearly distinguish the bright multi-line spectrum 
of A from a faint spectrum for B that is dominated by a light spot at $\sim$ 4760 
\AA. In order to extract the spectra of A and B, we followed a simple procedure. 
Our 1D model at each wavelength bin incorporated two Gaussian profiles with a 
fixed separation between both sources. In a first iteration, there were four free 
parameters: the position of A, the width of the Gaussian function, and the total 
fluxes of A and B. We then fitted A positions and Gaussian widths to smooth 
polynomial functions of the observed wavelength. In a second iteration, only the 
total fluxes were allowed to vary. A very similar procedure is followed in 
Sec.~\ref{sec:q1617}. 

\begin{table}
\centering
\caption{NOT spectra of the two main components in SDSS J1642+3200. Observed 
wavelengths ($\lambda$ values) are in \AA\ and fluxes ($F_{\lambda}$ values) are 
in 10$^{-17}$ erg cm$^{-2}$ s$^{-1}$ \AA$^{-1}$. The full table is available 
online in a machine-readable ASCII format. A portion is shown here for guidance 
regarding its form and content.}
\label{tab:q1642dat}
\begin{tabular}{ccc} 
\hline
$\lambda$ & $F_{\lambda}$(A) & $F_{\lambda}$(B) \\
\hline
3576.500 & 44.397 & 15.231 \\
3577.323 & 37.889 &  4.632 \\
3578.146 & 31.381 &  4.183 \\
3578.969 & 29.859 & 14.035 \\
3579.792 & 26.101 & 15.786 \\
3580.615 & 29.631 & 14.101 \\
3581.439 & 36.411 & 12.415 \\
3582.262 & 43.191 &  9.497 \\
\hline
\end{tabular}
\end{table}

\begin{figure}
\centering
\includegraphics[width=\columnwidth]{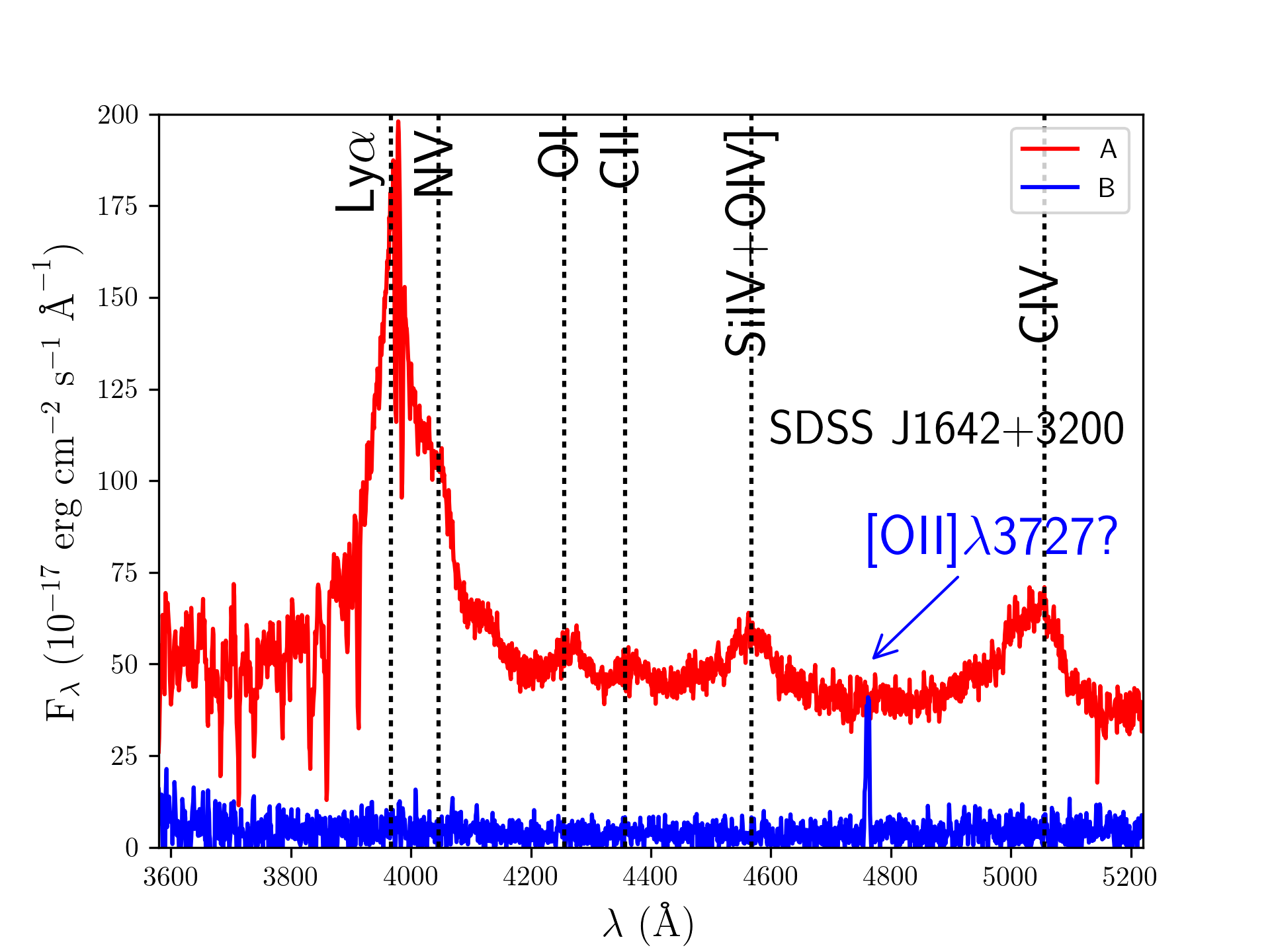}
\caption{NOT spectra of SDSS J1642+3200. Here, A (red line) is the confirmed 
quasar and B (blue line) its companion. Vertical dotted lines indicate emission 
lines at $z$ = 2.263.}
\label{fig:NOTq1642AB}
\end{figure}

The final flux-calibrated spectra of A and B are included in 
Table~\ref{tab:q1642dat} and depicted in Fig.~\ref{fig:NOTq1642AB}. While A is a 
quasar with Ly$\alpha$, \ion{N}{v}, \ion{O}{i}, \ion{C}{ii}, 
\ion{Si}{iv}/\ion{O}{iv}] and \ion{C}{iv} emission lines at the expected redshift 
of 2.263, the spectral energy distribution of B reveals a weak flat continuum 
with a superimposed narrow emission line. This prominent emission may correspond 
to the \ion{O}{ii} $\lambda\lambda$ 3727 line at $z$ = 0.277, since such redshift 
is within the error bar for the photometric $z$ of B+R in the SDSS database. 
Hence, B is not the second image of the quasar at $z$ = 2.263, but another active 
galactic nucleus showing variability, a flat continuum and narrow emission lines 
at optical wavelengths. The residual source R is most likely the host galaxy of 
B. 

For SDSS J1642+3200, the question arises why the host galaxy of B does not 
produce strongly lensed images of the background source associated with A. 
To answer this question, we 
considered a simple model for the mass distribution in such host galaxy. The 
matter content of the deflector was modelled as a singular isothermal sphere 
(SIS), so its Einstein radius is given by $\theta_{\rm E}$ = 1\farcs4 
($\sigma_{\rm v}$/220 km s$^{-1})^2 (D_{\rm ds}/D_{\rm s})$, where $\sigma_{\rm 
v}$ is the one-dimensional velocity dispersion, $D_{\rm ds}$ is the (angular 
diameter) distance between deflector and source, and $D_{\rm s}$ is the distance 
between observer and source \citep[e.g.][]{Nar99}. Using suitable expressions for 
these distances \citep[e.g.][]{Hog99}, a concordance flat cosmology \citep{Ben14}
and the redshifts involved: $z_{\rm d}$ = 0.277 and $z_{\rm s}$ = 2.263, the SIS 
model yielded $\theta_{\rm E}$ = 1\farcs12 ($\sigma_{\rm v}$/220 km s$^{-1})^2$. 
To see only one image of the distant quasar, the condition that has to be 
fulfilled is $\Delta \theta > 2\theta_{\rm E}$ (weak lensing regime), or 
equivalently, the deflecting mass within 10 kpc must be less than 3 $\times$ 
10$^{11}$ M$_{\odot}$ ($\sigma_{\rm v} <$ 255 km s$^{-1}$).   

\section{SDSS J2153+2732: a puzzling candidate}
\label{sec:q2153}

Besides the three golden candidates in \citetalias{Ser16} (see also the first 
paragraph of Sec.~\ref{sec:intro}), we found a puzzling candidate in the SS2 
subsample. Although there is no evidence for a lensing galaxy (residual source) 
in SDSS J2153+2732, this system displayed significant anti-parallel variations in 
the flux of its two components A and B. The brighter component A is classified as 
a quasar at $z$ = 2.214 with $r \sim$ 19, whereas B is about one magnitude 
fainter (see the bottom left panel of Fig.~\ref{fig:acquis}). The SDSS 
position of A is RA (J2000) = 328\fdg31784 and Dec. (J2000) = +27\fdg54302, and 
the two sources are separated by 3\farcs6. We observed SDSS J2153+2732 on 2015 
November 7 using the ALFOSC spectrograph (grism \#14) on the NOT. The 
1\farcs0-width slit was aligned with the axis defined by A and B, and we took a 
global exposure of 4000 s with FWHM seeing of about 1\arcsec. We obtained the 
spectra of both components following the same method as in Sec.~\ref{sec:q1642}. 

\begin{figure}
\includegraphics[width=\columnwidth]{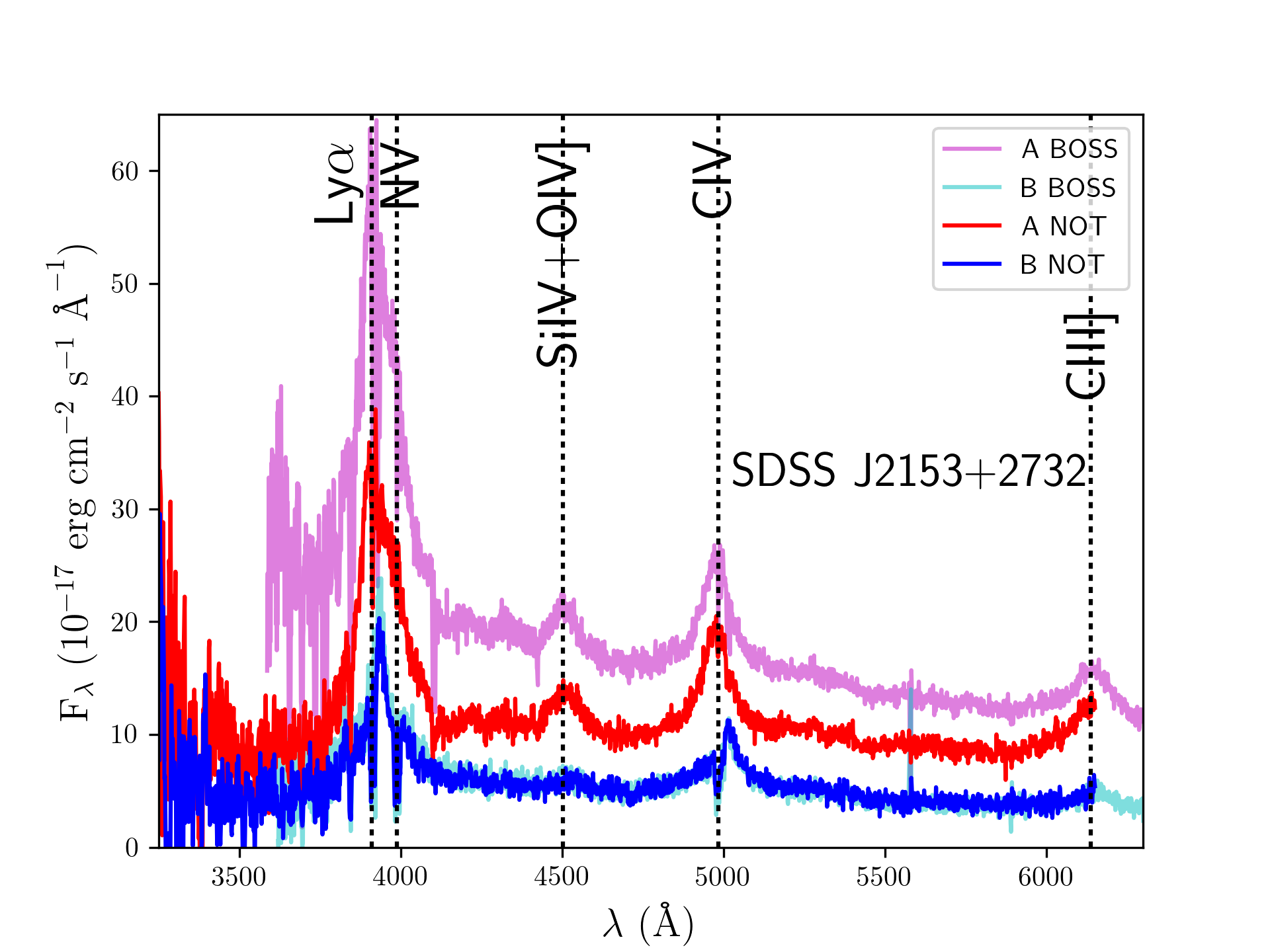}
\includegraphics[width=\columnwidth]{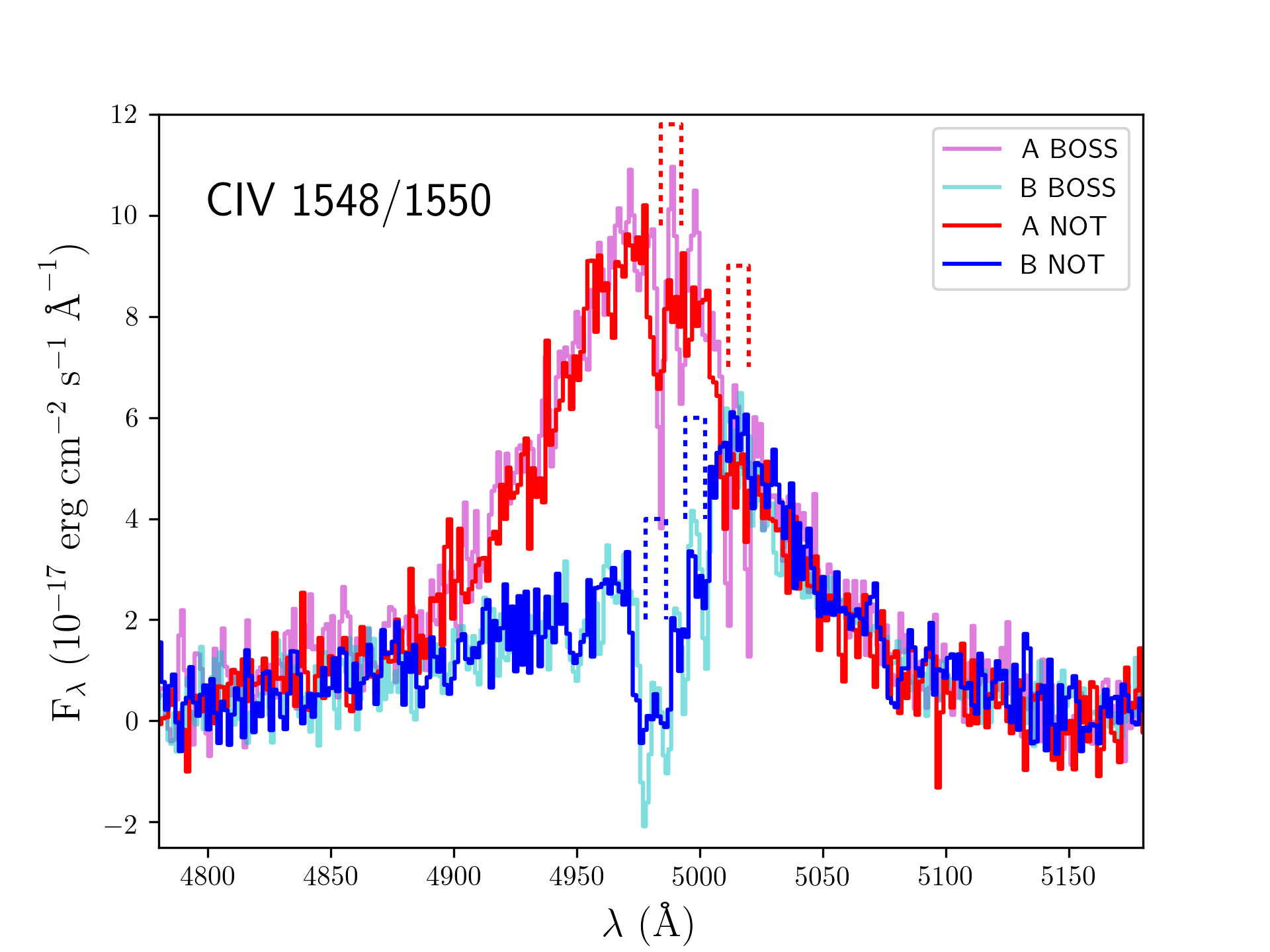}
\caption{Spectra of SDSS J2153+2732. Top: NOT spectra of the confirmed quasar A 
(red line) and its companion B (blue line) in 2015 November. We also display the 
early SDSS-BOSS spectrum of A in 2012 June (magenta line) and the almost 
contemporary SDSS-BOSS spectrum of B in 2015 October (cyan line). Vertical dotted 
lines indicate emission lines at $z$ = 2.214. Bottom: \ion{C}{iv} 
$\lambda\lambda$ 1548, 1550 absorption doublets distorting the \ion{C}{iv} 
emission-line profiles. We use dotted top-hat marks to highlight the positions of
these doublets.}
\label{fig:NOTq2153AB}
\end{figure}

Our final results are shown in Table~\ref{tab:q2153dat} and 
Fig.~\ref{fig:NOTq2153AB}. In the top panel of Fig.~\ref{fig:NOTq2153AB} (red 
and blue lines), it is evident that A and B are two different quasars having 
similar (but not identical) redshifts. The SDSS-IV DR14 \citep{Abo18} also 
includes the BOSS spectrum of B taken on 2015 October 12, only a few weeks before 
the NOT observations and $\sim$ 3 years after the SDSS-BOSS spectrum of A 
obtained on 2012 June 20. In agreement with our NOT spectra, the current version 
of the SDSS database indicates that SDSS J2153+2732 consists of a pair of quasars 
with $z$(A) = 2.214 and $z$(B) = 2.244. 

\begin{table}
\centering
\caption{NOT spectra of the two components in SDSS J2153+2732. Observed 
wavelengths ($\lambda$ values) are in \AA\ and fluxes ($F_{\lambda}$ values) are 
in 10$^{-17}$ erg cm$^{-2}$ s$^{-1}$ \AA$^{-1}$. The full table is available 
online in a machine-readable ASCII format. A portion is shown here for guidance 
regarding its form and content.}
\label{tab:q2153dat}
\begin{tabular}{ccc} 
\hline
$\lambda$ & $F_{\lambda}$(A) & $F_{\lambda}$(B) \\
\hline
3245.585 & -12.216 & -43.301 \\
3247.004 &  14.016 & -21.166 \\
3248.422 &  40.247 &   7.852 \\
3249.841 &  36.725 &   7.782 \\
3251.259 &  27.255 &   7.713 \\
3252.678 &  33.396 &  29.561 \\
3254.096 &  32.233 &  25.537 \\
3255.515 &  31.071 &  24.027 \\
\hline
\end{tabular}
\end{table}

We analysed the SDSS-BOSS spectra of both sources (see the magenta and cyan lines 
in the top panel of Fig.~\ref{fig:NOTq2153AB}; we show the spectral region 
covering wavelengths shorter than 6300 \AA) to estimate reliable redshifts and 
discuss whether or not the two quasars are physically associated. Although the 
\ion{C}{iii}] and \ion{C}{iv} emission lines of A led to an average redshift of 
2.214, the \ion{Mg}{ii} emission line of A yielded a larger value of 2.231. This 
last emission line is thought to be a good tracer of the systemic redshift 
\citep[e.g.][]{Ric02}, and we adopted $z$(A) = 2.231 instead of the nominal value 
in the SDSS database. For the other quasar, the \ion{Mg}{ii} emission line 
indicated that $z$(B) is 2.242, which is very close to its nominal value. 

As close galaxy encounters trigger or enhance their nuclear activity 
\citep[e.g.][]{Mor99,Hen06a,Eft17}, we checked the hypothesis of a close
separation quasar-quasar pair. Thus, assuming a cosmological redshift $z = 
\left[z({\rm A}) + z({\rm B})\right]/2 \sim$ 2.237 for the two quasars, the 
physical transverse separation and the peculiar velocity difference (along the 
line-of-sight) would be $\sim$ 30 kpc\footnote{Based on the angular separation 
$\Delta \theta$ = 3\farcs6, the concordance flat model described by 
\citet{Ben14} and the cosmology calculator of \citet{Wri06}} and $\sim$ 1000 km 
s$^{-1}$, respectively. Even though such large relative motion requires very 
massive host galaxies in a binary scenario, quasars likely reside in central 
regions of massive halos \citep[e.g.][]{Tur91,Por04,Pro13}. Moreover, considering 
a collision or merger in a high density environment \citep[e.g.][and references 
therein]{Hen06a}, it seems also plausible to account for the estimated velocity 
difference.   

When a \ion{C}{iv} absorption system is found in a sightline towards a quasar, 
there is a high probability to detect \ion{C}{iv} absorption within 1000 km 
s$^{-1}$ in a close sightline \citep[e.g.][]{Mar10}. \citet{Pro14} also reported
that quasars at $z \sim$ 2 are embedded in massive dark matter halos. In 
addition, quasar sightlines often show \ion{C}{iv} absorption, and most absorbers 
may be associated to massive galaxies that cluster with the quasar hosts. We 
searched through the SDSS-BOSS spectra for \ion{C}{iv} absorbers in the 
surroundings of the two quasars, detecting several \ion{C}{iv} $\lambda\lambda$ 
1548, 1550 absorption doublets. The sightline towards A exhibits absorption at 
three different redshifts: $z_{\rm A1}$ = 2.2369, $z_{\rm A2}$ = 2.2193 and 
$z_{\rm A3}$ = 1.6480, whereas the other sightline only shows the presence of 
highly ionised \ion{C}{iv} gas at $z_{\rm B1}$ = 2.2256 and $z_{\rm B2}$ = 
2.2153. The four doublets at $z >$ 2 are illustrated in the bottom panel of 
Fig.~\ref{fig:NOTq2153AB}. We also found \ion{H}{i} Ly$\alpha$ absorbers at 
redshifts similar to those of the \ion{C}{iv} systems at $z >$ 2. Additionally, 
we identified \ion{N}{v} gas linked to A2 and B2. 

Considering a systemic redshift $z$(A) = 2.231 (see here above), the absorber A1 
is moving towards the quasar at a speed of $\sim$ 550 km s$^{-1}$. Alternatively, 
the systemic redshift of A might be slightly underestimated, so $z({\rm A}) \sim 
z_{\rm A1}$ and the peculiar velocity difference (between A and B) would be less 
than 500 km s$^{-1}$. This clearly favours a close galaxy encounter. It is also 
reasonable to think that the galaxies harbouring the four intervening systems at 
2.215 $< z <$ 2.237 are physically associated with each other and with the two 
quasar host galaxies. Therefore, despite a binary scenario cannot be ruled out 
from current data, a collision/merger in a high density region is supported by 
evidence of clustering around SDSS J2153+2732.

\section{The pseudo-galaxy SDSS J0240$-$0208}
\label{sec:q0240}

Searching for pairs of point-like sources with small angular separation, we  
found the system SDSS J0240$-$0208 at RA (J2000) = 40\fdg07669 and Dec. (J2000) = 
$-$2\fdg14730. This consists of two objects A and B ($r \sim$ 18$-$19) that are 
separated by 0\farcs95 (see the bottom right panel of 
Fig.~\ref{fig:acquis}). We performed follow-up spectroscopic observations of 
SDSS J0240$-$0208 on 2015 November 5 under good seeing conditions and using the 
NOT/ALFOSC 1\farcs0-width slit. The global exposure of 3000 s covered a sky area 
including both sources and the 3440$-$6150 \AA\ wavelength range (grism \#14). 
From these new data and the spectrum extraction techniques in previous sections, 
we then obtained the spectral energy distributions of A and B. The pair is 
misidentified as a galaxy in the SDSS-IV DR14 \citep{Abo18}, and there are no 
available spectra in the SDSS database. 

\begin{figure}
\centering
\includegraphics[width=\columnwidth]{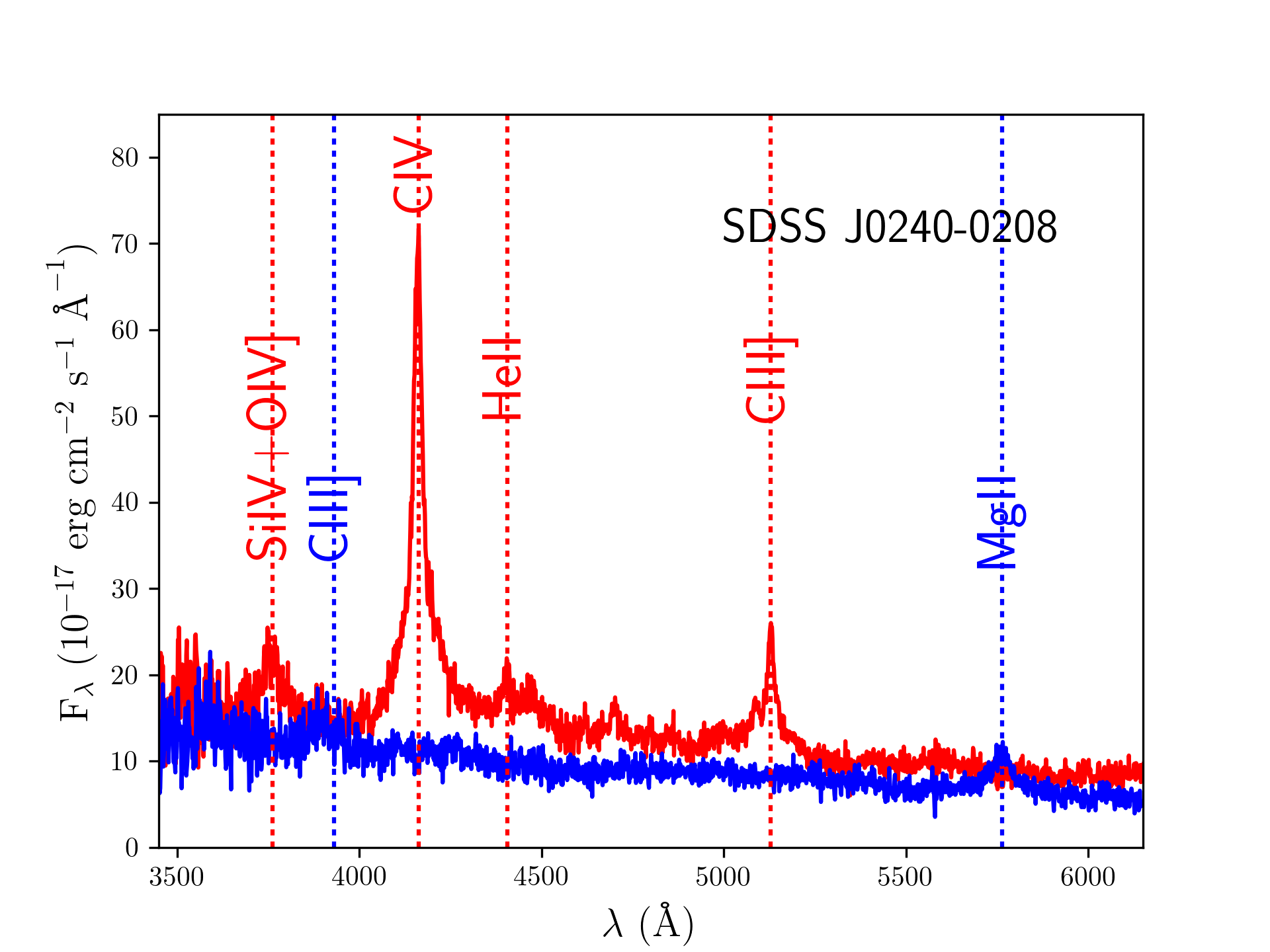}
\caption{NOT spectra of SDSS J0240$-$0208. Here, A (red line) and B (blue line) 
are two previously unknown quasars. Vertical dotted lines indicate emission lines 
at $z$ = 1.687 (red) and $z$ = 1.059 (blue).}
\label{fig:NOTq0240AB}
\end{figure}

The NOT spectra of A and B are listed in Table~\ref{tab:q0240dat} and shown in 
Fig.~\ref{fig:NOTq0240AB}. The spectrum of the brighter component A (red line in 
Fig.~\ref{fig:NOTq0240AB}) contains \ion{Si}{iv}/\ion{O}{iv}], \ion{C}{iv}, 
\ion{He}{ii} and \ion{C}{iii}] emissions at $z$ = 1.687. However, the spectrum of 
the fainter component B (blue line in Fig.~\ref{fig:NOTq0240AB}) only includes a 
clear emission feature at 5765 \AA. Such feature is probably produced by 
\ion{Mg}{ii} gas at $z$ = 1.059, which would explain the flux enhancement near 
3900 \AA\ as due to \ion{C}{iii}] emission. The putative redshift of 1.059 was 
also checked by analysing the continuum of B. This spectral contribution can be 
easily reproduced by adding the iron forest generated by a quasar at $z$ = 1.059 
\citep[we used a scaled version of the Fe pseudo-continuum template of][]{Ves01} 
to a nuclear power-law continuum. Thus, SDSS J0240$-$0208 is most likely a 
projected quasar pair, i.e., a pair of quasars that appear to be very close to 
each other on the sky but are in fact at a large distance from each other in 
space.

\begin{table}
\centering
\caption{NOT spectra of the two components in SDSS J0240$-$0208. Observed 
wavelengths ($\lambda$ values) are in \AA\ and fluxes ($F_{\lambda}$ values) are 
in 10$^{-17}$ erg cm$^{-2}$ s$^{-1}$ \AA$^{-1}$. The full table is available 
online in a machine-readable ASCII format. A portion is shown here for guidance 
regarding its form and content.}
\label{tab:q0240dat}
\begin{tabular}{ccc} 
\hline
$\lambda$ & $F_{\lambda}$(A) & $F_{\lambda}$(B) \\
\hline
3438.344 & 21.476 & 10.413 \\
3439.773 & 20.684 &  9.872 \\
3441.202 & 18.984 &  9.064 \\
3442.632 & 15.626 & 12.295 \\
3444.061 & 15.795 &  9.951 \\
3445.491 & 15.964 &  7.606 \\
3446.920 & 24.354 & 11.958 \\
3448.349 & 20.607 & 13.959 \\
\hline
\end{tabular}
\end{table}

To obtain an upper limit on the mass of the host galaxy of B, we followed a 
procedure identical to the one used in the last paragraph of 
Sec.~\ref{sec:q1642}. For SDSS J0240$-$0208, $z_{\rm d}$ = 1.059, $z_{\rm s}$ = 
1.687 and $\theta_{\rm E}$ = 0\farcs38 ($\sigma_{\rm v}$/220 km s$^{-1})^2$. 
Hence, the $\Delta \theta > 2\theta_{\rm E}$ relationship leads to $M_{\rm host} 
<$ 2.9 $\times$ 10$^{11}$ M$_{\odot}$ ($\sigma_{\rm v} <$ 250 km s$^{-1}$) within 
a radius of 10 kpc. We are dealing with a weak lensing effect and a single image 
A located $\Delta \theta \sim$ 1\arcsec\ from B. Assuming a characteristic 
velocity dispersion of $\sigma_{\rm v} \sim$ 225 km s$^{-1}$ \citep[e.g.][and 
references therein]{Rus03}, the background quasar and B would be separated by 
$\beta = \Delta \theta - \theta_{\rm E} \sim$ 0\farcs6, and A would be magnified 
by a factor $\mu_{\rm A} = 1 + \theta_{\rm E}/\beta \sim$ 1.7 
\citep[e.g.][]{Nar99}. This characteristic host galaxy can only produce two 
images of a $z$ = 1.687 quasar when the source lies within $\theta_{\rm E} \sim$ 
0\farcs4 of B (strong lensing regime).

\section{Summary}
\label{sec:end}

The search for new gravitational lensed quasars is a task of great importance in 
cosmology \citep[e.g.][]{Sch06}, and thus several data mining methods have been 
used to select lensed quasar candidates in large databases. As quasars are 
variable sources, some selection techniques based on variability analyses were 
proposed over the previous decade \citep[e.g.][]{Pin05,Koc06,Lac09}, and a number 
of multiple quasars are being discovered in recent years from methods including 
variability studies \citep[e.g.][]{Ser16,Ber17,Kos18}. In \citetalias{Ser16}, we 
selected three superb candidates for double quasar: SDSS J1442+4055, SDSS 
J1617+3827 and SDSS J1642+3200. Each of these golden targets consisted of a 
confirmed quasar, a neighbour point-like source without spectroscopic 
identification and an extended residual source. Additionally, as expected for 
double quasars, the two point-like objects showed significant parallel flux 
variations on a long timescale. 

We initially reported the discovery of the optically bright, wide separation 
double quasar SDSS J1442+4055 \citepalias{Ser16}, and new GTC data indicate that 
SDSS J1617+3827 is most likely another lensed quasar at $z$ = 2.079 (this paper). 
The GTC spectra of both images of SDSS J1617+3827 exhibit identical emission-line 
redshifts, and reveal the presence of two intervening objects at redshifts of 
0.602 and 0.866, with the nearest being an early-type (secondary) lensing galaxy. 
Moreover, our GTC acquisition frame unveils the presence of a candidate for the 
main lensing galaxy, located 0\farcs4 from the faintest quasar image. In this 
paper, using new spectroscopic observations with the NOT, we also identify SDSS 
J1642+3200 as a projected pair of active galactic nuclei. Thus, in comparison 
with other standard selection techniques that are largely inefficient 
\citep[e.g.][]{Ina12}, 66\% of our superb candidates turn out to be double 
quasars. A suitable selection of golden targets leads to an impressive efficiency 
in discovering multiple quasars, saving large amounts of spectroscopic observing 
time.

The spectroscopic identification campaign with the NOT has been also proved to be 
a useful way to find pairs of active galactic nuclei. In addition to SDSS J1642+3200,
we report on the projected quasar pair SDSS J0240$-$0208, which consists of two
distinct quasars at $z$ = 1.059 and $z$ = 1.687. Using astrometry and 
spectroscopy (redshifts) of such pairs, as well as a SIS mass model for the 
foreground host galaxies and a concordance cosmology, we determine the constraint 
$\sigma_{\rm v} <$ 250$-$255 km s$^{-1}$ for the one-dimensional velocity 
dispersion. Therefore, our results are consistent with typical dark matter halos
having $\sigma_{\rm v} \sim$ 225 km s$^{-1}$ \citep{Rus03}. Regarding the third 
NOT target, SDSS J2153+2732 is a quasar pair inhabiting the central regions of 
two close galaxies at $z \sim$ 2.24. These findings are supported by the new NOT
spectra and updated information in the SDSS database. The SDSS-BOSS spectra of 
both quasars show a series of \ion{C}{iv} absorbers at $z >$ 2 that could be
associated with massive galaxies close to the quasar hosts \citep[high density 
environment; e.g.][]{Pro14}.   
 
\section*{Acknowledgements}

We thank the anonymous referee for helpful comments that contributed to improving 
the final version of the paper.
This paper is based on observations made with the Gran Telescopio Canarias (Prog. 
GTC69-17A), the Liverpool Telescope (Prog. XCL04BL2) and the Nordic Optical 
Telescope (Progs. SST2016-318 and 52-402), operated on the island of La Palma by 
GRANTECAN S.A., the Liverpool John Moores University (with financial support from 
the UK Science and Technology Facilities Council) and the Nordic Optical 
Telescope Scientific Association, respectively, in the Spanish Observatorio del 
Roque de los Muchachos of the Instituto de Astrof\'{\i}sica de Canarias. We thank 
the staff of the three telescopes for a kind interaction before, during and after 
the observations. We also used data taken from the Sloan Digital Sky Survey 
(SDSS) web sites, and we are grateful to the SDSS collaboration for doing that 
public database. We gratefully acknowledge to Sh. Ehgamberdiev and to staff of 
Maidanak Observatory of Astronomical Institute UzAS for supporting our 
observations at Maidanak Astronomical Observatory. This research has been 
supported by the Spanish Department of Research, Development and Innovation grant 
AYA2013-47744-C3-2-P, the GLENDAMA (AYA2017-89815-P) project financed by 
MINECO/AEI/FEDER-UE, the complementary action "Lentes Gravitatorias y Materia 
Oscura" financed by the SOciedad para el DEsarrollo Regional de CANtabria 
(SODERCAN S.A.) and the Operational Programme of FEDER-UE, and the University of 
Cantabria.

%%%%%%%%%%%%%%%%%%%% REFERENCES %%%%%%%%%%%%%%%%%%
% The best way to enter references is to use BibTeX:

%\bibliographystyle{mnras}
%\bibliography{example} % if your bibtex file is called example.bib

\begin{thebibliography}{999}
\bibitem[\protect\citeauthoryear{Abolfathi et al.}{2018}]{Abo18}
Abolfathi B. et al., 2018, \apjs, 235, 42 
\bibitem[\protect\citeauthoryear{Ahn et al.}{2014}]{Ahn14}
Ahn C. P. et al., 2014, \apjs, 211, 17 
\bibitem[\protect\citeauthoryear{Bennett et al.}{2014}]{Ben14}
Bennett C. L., Larson D., Weiland J. L., Hinshaw G., 2014, \apj, 794, 135 
\bibitem[\protect\citeauthoryear{Berghea et al.}{2017}]{Ber17} 
Berghea C. T., Nelson G. J., Rusu C. E., Keeton C. R., Dudik R. P., 2017, 
\apj, 844, 90
\bibitem[\protect\citeauthoryear{Claeskens et al.}{2000}]{Cla00} 
Claeskens J. F., Lee D. W., Remy M., Sluse D., Surdej J., 2000, \aap, 356, 840
\bibitem[\protect\citeauthoryear{Djorgovski et al.}{1987}]{Djo87}
Djorgovski S., Perley R., Meylan G., McCarthy P., 1987, \apj, 321, L17 
\bibitem[\protect\citeauthoryear{Eftekharzadeh et al.}{2017}]{Eft17}
Eftekharzadeh S. et al., 2017, \mnras, 468, 77 
\bibitem[\protect\citeauthoryear{Filippenko}{1982}]{Fil82} 
Filippenko A. V., 1982, \pasp, 94, 715
\bibitem[\protect\citeauthoryear{Goicoechea \& Shalyapin}{2016}]{Goi16}
Goicoechea L. J., Shalyapin V. N., 2016, \aap, 596, 77 
\bibitem[\protect\citeauthoryear{Hennawi et al.}{2006a}]{Hen06a}
Hennawi J. F. et al., 2006a, \aj, 131, 1 
\bibitem[\protect\citeauthoryear{Hennawi et al.}{2006b}]{Hen06b}
Hennawi J. F. et al., 2006b, \apj, 651, 61 
\bibitem[\protect\citeauthoryear{Hogg}{1999}]{Hog99} 
Hogg D. W., 1999, eprint arXiv:astro-ph/9905116
\bibitem[\protect\citeauthoryear{Inada et al.}{2012}]{Ina12}
Inada N. et al., 2012, \aj, 143, 119 
\bibitem[\protect\citeauthoryear{Jackson et al.}{2012}]{Jac12} 
Jackson N., Rampadarath H., Ofek E. O., Oguri, M., Shin, M., 2012, \mnras, 
419, 2014 
\bibitem[\protect\citeauthoryear{Kayo \& Oguri}{2012}]{Kay12} 
Kayo I., Oguri, M., 2012, \mnras, 424, 1363 
\bibitem[\protect\citeauthoryear{Kochanek et al.}{2006}]{Koc06} 
Kochanek C. S., Mochejska B., Morgan N. D., Stanek K. Z., 2006, \apj, 637, L73 
\bibitem[\protect\citeauthoryear{Kostrzewa-Rutkowska et al.}{2018}]{Kos18} 
Kostrzewa-Rutkowska Z. et al., 2018, \mnras, 476, 663 
\bibitem[\protect\citeauthoryear{Lacki et al.}{2009}]{Lac09} 
Lacki B. C., Kochanek C. S., Stanek K. Z., Inada, N., Oguri, M., 2009, \apj, 
698, 428 
\bibitem[\protect\citeauthoryear{Lau et al.}{2018}]{Lau18} 
Lau M. W., Prochaska J. X., Hennawi J. F., 2018, \apj, 857, 126 
\bibitem[\protect\citeauthoryear{Lawther et al.}{2012}]{Law12}
Lawther D. et al., 2012, \aap, 546, 67
\bibitem[\protect\citeauthoryear{Martin et al.}{2010}]{Mar10}
Martin C. L. et al., 2010, \apj, 721, 174 
\bibitem[\protect\citeauthoryear{More et al.}{2016}]{Mor16}
More A. et al., 2016, \mnras, 456, 1595 
\bibitem[\protect\citeauthoryear{Mortlock et al.}{1999}]{Mor99}
Mortlock D. J., Webster R. L., Francis P. J., 1999, \mnras, 309, 836 
\bibitem[\protect\citeauthoryear{Narayan \& Bartelmann}{1999}]{Nar99}
Narayan R., Bartelmann M., 1999, Lectures on Gravitational Lensing, in 
Dekel A., Ostriker J.P., eds, Formation of Structure in the Universe. 
Cambridge University Press, Cambridge (eprint arXiv:astro-ph/9606001)
\bibitem[\protect\citeauthoryear{P\^aris et al.}{2014}]{Par14}
P\^aris I. et al., 2014, \aap, 563, 54
\bibitem[\protect\citeauthoryear{Peng et al.}{2002}]{Pen02}
Peng C. Y., Ho L. C., Impey C. D., Rix H.-W., 2002, \aj, 124, 266
\bibitem[\protect\citeauthoryear{Peng et al.}{2010}]{Pen10}
Peng C. Y., Ho L. C., Impey C. D., Rix H.-W., 2010, \aj, 139, 2097
\bibitem[\protect\citeauthoryear{Pindor}{2005}]{Pin05}
Pindor B., 2005, \apj, 626, 649 
\bibitem[\protect\citeauthoryear{Porciani et al.}{2004}]{Por04}
Porciani C., Magliocchetti M., Norberg P., 2004, \mnras, 355, 1010 
\bibitem[\protect\citeauthoryear{Prochaska et al.}{2013}]{Pro13}
Prochaska J. X. et al., 2013, \apj, 776, 136 
\bibitem[\protect\citeauthoryear{Prochaska et al.}{2014}]{Pro14}
Prochaska J. X., Lau M. W., Hennawi J. F., 2014, \apj, 796, 140 
\bibitem[\protect\citeauthoryear{Prochter et al.}{2006}]{Pro06}
Prochter G. E., Prochaska J. X., Burles S. M., 2006, \apj, 639, 766 
\bibitem[\protect\citeauthoryear{Richards et al.}{2002}]{Ric02}
Richards G. T. et al., 2002, \aj, 124, 1 
\bibitem[\protect\citeauthoryear{Rusin et al.}{2003}]{Rus03}
Rusin D. et al., 2003, \apj, 587, 143
\bibitem[\protect\citeauthoryear{Schneider, Kochanek \& Wambsganss}{2006}]{Sch06}
Schneider P., Kochanek C. S., Wambsganss J., 2006, in Meylan G., Jetzer P., 
North P., eds, Proc. 33rd Saas-Fee Advanced Course, Gravitational Lensing: 
Strong, Weak \& Micro. Springer, Berlin
\bibitem[\protect\citeauthoryear{Sergeyev et al.}{2016}]{Ser16}
Sergeyev A. V., Zheleznyak A. P., Shalyapin V. N., Goicoechea L. J., 2016, 
\mnras, 456, 1948  
\bibitem[\protect\citeauthoryear{Seyffert et al.}{2013}]{Sey13}
Seyffert E. N. et al., 2013, \apj, 779, 161 
\bibitem[\protect\citeauthoryear{Shalyapin \& Goicoechea}{2014}]{Sha14}
Shalyapin V. N., Goicoechea L. J., 2014, \aap, 568, 116
\bibitem[\protect\citeauthoryear{Treu}{2010}]{Tre10} 
Treu T., 2010, \araa, 48, 87
\bibitem[\protect\citeauthoryear{Turner}{1991}]{Tur91} 
Turner E. L., 1991, \aj, 101, 5
\bibitem[\protect\citeauthoryear{Vestergaard \& Wilkes}{2001}]{Ves01} 
Vestergaard M., Wilkes B.~J., 2001, \apjs, 134, 1
\bibitem[\protect\citeauthoryear{Wright}{2006}]{Wri06} 
Wright E. L., 2006, \pasp, 118, 1711
\end{thebibliography}

% Alternatively you could enter them by hand, like this:
% This method is tedious and prone to error if you have lots of references

%%%%%%%%%%%%%%%%%%%%%%%%%%%%%%%%%%%%%%%%%%%%%%%%%%

% Don't change these lines
\bsp	% typesetting comment
\label{lastpage}
\end{document}